\documentclass[superscriptaddress,twocolumn,preprintnumbers,amsmath,amssymb]{revtex4}
\usepackage{graphicx}% Include figure files
\usepackage{dcolumn}% Align table columns on decimal point
\usepackage{bm}% bold math

\begin{document}

%\preprint{APS/123-QED}

\title{Highly efficient 4 micron light generation through fs-fiber laser driven supercontinuum in Si\textsubscript{3}N\textsubscript{4} waveguides}% Force line breaks with \\

\author{Davide~Grassani}
 \email{davide.grassani@epfl.ch}
\author{ Eirini~Tagkoudi}
\affiliation{Photonics Systems Laboratory (PHOSL), {\'E}cole Polytechnique F{\'e}d{\'e}rale de Lausanne (EPFL), CH-1015 Lausanne, Switzerland \\}

\author{ Hairun~Guo}
\affiliation{Laboratory of Photonics and Quantum Measurements (LPQM), {\'E}cole Polytechnique F{\'e}d{\'e}rale de Lausanne (EPFL), CH-1015 Lausanne, Switzerland \\}

\author{ Clemens~Herkommer} 
\affiliation{Laboratory of Photonics and Quantum Measurements (LPQM), {\'E}cole Polytechnique F{\'e}d{\'e}rale de Lausanne (EPFL), CH-1015 Lausanne, Switzerland \\}
\affiliation{Physik-Department, Technische Universit{\"a}t M{\"u}nchen (TUM), D-80333 M{\"u}nchen, Germany}

\author{Tobias~J.~Kippenberg} 
\affiliation{Laboratory of Photonics and Quantum Measurements (LPQM), {\'E}cole Polytechnique F{\'e}d{\'e}rale de Lausanne (EPFL), CH-1015 Lausanne, Switzerland \\}

\author{Camille-Sophie~Br{\`e}s} 
\affiliation{Photonics Systems Laboratory (PHOSL), {\'E}cole Polytechnique F{\'e}d{\'e}rale de Lausanne (EPFL), CH-1015 Lausanne, Switzerland \\}


\begin{abstract}
Directly accessing the middle infrared, the molecular functional group spectral region, via supercontinuum generation processes based on turn-key fiber lasers offers the undeniable advantage of simplicity and robustness. Recently, the assessment of the coherence of the mid-IR dispersive wave in silicon nitride waveguides, pumped at telecom wavelength, established an important first step towards mid-IR frequency comb generation based on such compact systems. Yet, the spectral reach and efficiency still fall short for practical implementation. Here, we experimentally demonstrate for the first time to our knowledge, that fs-fiber laser driven systems based on large-cross section silicon nitride waveguides can reach, with powers sufficient to drive dual-comb spectroscopy, the important greenhouse gases spectral region near 4 micron, typically accessed through different frequency generation or more complex approaches. We show, from a 2 $\mu$m femtosecond fiber laser, up to 30\% power conversion and milliwatt-level output powers, proving that such sources are suitable candidate for compact, chip-integrated spectroscopic and sensing applications.
\end{abstract}

%Use showkeys class option if keyword

\maketitle

\section{\label{sec:level1}introduction}

The middle infrared (mid-IR) spectral region (2-10  $\mu$m) has a high technological importance for spectroscopy and sensing with applications in health and environmental monitoring \cite{Keilmann2004,Petersen2014}.  Particularly, the 3 – 5 $\mu$m region, the functional group region, is interesting as it hosts the first mid-IR atmospheric window, and  contains the molecular fingerprint of many hydrocarbons, nitrogen dioxide, and specimens detectable in breath analysis \cite{Vainio2016}. Beside direct mid-IR generation with Quantum Cascade Lase\cite{Razeghi2013}, Interband Cascade Lasers \cite{Vurgaftman2009} and Fe2+ doped crystals \cite{Kozlovskii2011}, wavelength conversion in nonlinear materials pumped by fiber lasers is a promising solution to reach the mid-IR spectral range.  In fact, femtosecond mode-locked fiber lasers are a reliable, easy-to-use and compact platform and their spectra provide a frequency comb which can enable more advanced spectroscopic techniques with improved resolution, bandwidth and sensitivity \cite{Bernhardt2010,Diddams2007,Suh2016}. Yet, to date accessing the mid-IR spectral region with femtosecond fiber lasers is challenging, and typically only achieved with difference frequency generation schemes \cite{Alvarez2008,Ruehl2012,Zhu2013,Yao2012,Erny2007,Cruz2015a,Sobon2017,Ycas2018}. 

In the last years, the use of suitable rare earth dopant such as Thulium (Tm) and Holmium (Ho), together with nonlinear frequency shift by intrapulse Raman scattering, enabled to red shift the emission wavelength of femtosecond fiber lasers to the short-wave infrared (SWIR) long edge, around 2.8 $\mu$m \cite{Duval2015}. Nevertheless, for the most common and technological mature silica fiber lasers, the emission wavelength limit is lower, as absorption by silica increases exponentially beyond 2.1 $\mu$m \cite{Xing2017}. 
Broadband mid-IR generation has been also achieved by supercontinuum generation (SCG) in photonic integrated waveguides or optical fibers. The most common platforms are chalcogenide glasses (ChG)\cite{Hudson2017,Lee2014,Marandi2012,Yu2013a}, fluoride fibers \cite{Salem2015a}, aluminium nitride (AlN) \cite{Hickstein2017} and complementary metal–oxide–semiconductor (CMOS) compatible materials such as silicon (Si) \cite{Singh2018,Kuyken2015}, silicon-rich nitride \cite{Liu2016a} and stochiometric silicon nitride (Si\textsubscript{3}N\textsubscript{4}) \cite{Porcel2017a,ChavezBoggio2016,Grassani2016}. The advantages of the SCG approach lies in the ease of operation and compactness: it is in fact a single pass geometry which does not require any additional seed laser and temporal synchronization. Moreover, such nonlinear platforms are usually compact and have low power consumption. When CMOS compatible materials are employed, they can also be fabricated with lithographic precision and high yield. Recently, a proof-of-principle experiment demonstrated the coherence of the mid-IR dispersive wave (DW) generated in Si\textsubscript{3}N\textsubscript{4} through SCG using an Er doped femtosecond fiber laser \cite{Guo2018}, assessing de facto the comb nature of the mid-IR supercontinuum portion. However, the wavelength reach and efficiency are still not suitable for applications. Indeed, reaching efficient wavelength conversion beyond 3 $\mu$m is still difficult when fiber pump lasers are used. On the one hand, larger is the spectral span to be covered, lower is the power transferred in the targeted region and thus the efficiency. On the other hand, materials with higher nonlinearities such as ChG and Si suffer from photo-darkening and two and three photons absorption in the 2 micron region, which limit the amount of power that can be coupled in the device. Moreover, the different nonlinear phenomena taking part in the SCG process convert part of the pump energy over unwanted spectral bands, further decreasing the conversion efficiency in the region of interest \cite{Dudley2010}. 

In this work, we overcome these hurdles and demonstrate a turn-key, efficient and compact mid-IR source with power levels sufficient for spectroscopy application.  We use the process of DW generation in dispersion engineered stoichiometric silicon nitride (Si\textsubscript{3}N\textsubscript{4}) waveguides to convert a commercial femtosecond fiber laser, emitting at 2090 nm, to the 3 - 4 $\mu$m wavelength region. This result has been achieved by leveraging an optimized waveguide geometry, which includes the fabrication of large cross section waveguides through the photonic damascene process \cite{Pfeiffer2016}. This enables the design of high mode confinement waveguides with a favorable dispersion for the generation of mid-IR DW when used in combination with the employed laser. By changing the waveguide width we tune the central DW wavelength with lithographic precision from 3.2 to beyond 4 $\mu$m. At the chip output, we are able to obtain more than 1 mW of average power just in the mid-IR DW up to an emission wavelength of 4 $\mu$m, corresponding to more than 6 mW on chip. Record conversion efficiency (CE), defined as on-chip generated Mid-IR DW power over on-chip coupled pump laser power, as high as 30\% is reached at 3.2 $\mu$m. Finally, a systematic experimental and numerical study of the generation of mid-IR DW allows us to identify different working regimes, which provide new information on the efficiency and dynamics of the DW generation process as a function of the pump power.

\section{MATERIALS AND METHODS}

\subsection{Mid-IR dispersive wave generation in Si\textsubscript{3}N\textsubscript{4} waveguides}

Propagation of sufficiently powerful femtosecond laser pulses in the anomalous group velocity dispersion (GVD) region of a nonlinear waveguide can induce high order soliton dynamics \cite{Dudley2010}, where the soliton number $N$ of the pulse is defined as $N^{2} = L_D/L_{NL}$. The dispersive length is given by $L_D = T_2/\mid \beta_2 \mid$, with $T$ the pulse duration and $\beta_2$ the GVD. The nonlinear length is $L_{NL} = 1/\gamma P$, with $P$ the pulse peak power and $\gamma$ being the waveguide’s effective nonlinearity \cite{Agrawal2013}. High order soliton dynamics leads to an initial spectral broadening, caused by self-phase modulation, and subsequent temporal compression which are proportional to $N$. During this stage, the soliton can be perturbed by high order dispersion (HOD) or nonlinear terms resulting in the fission in its fundamental components \cite{Agrawal2013,Dudley2006}. Among all the different solitons, the fundamental one has the highest peak power, shortest duration and largest spectral bandwidth. It propagates without dispersing with a positive wavenumber given by the nonlinear phase term ($\gamma P/2$). Close to the pump wavelength, in the anomalous GVD region ($\beta_2 < 0$), any linear wave propagates with negative wavenumber ($\beta_2 \omega^2$), therefore the soliton propagates without perturbations. However, far enough from the central wavelength, HOD terms come into play and the wavenumber of the linear waves can turn to positive values. At the compression point, due to its large bandwidth, the soliton can thus be perturbed shedding energy to linear dispersive waves spectrally shifted from the pump \cite{Akhmediev1995}. The DW generation process can occur at the frequency where the phase velocity of the soliton pulse equals the one of the linear wave, and is thus given by the phase matching condition \cite{Akhmediev1995}:
\begin{equation}
\beta(\omega) - \beta_(\omega_s) - v_{g}^{-1}(\omega - \omega_s) = \frac{\gamma P}{2}
\label{eq:one}
\end{equation}
where $\beta$ is the mode propagation vector, $\omega_s$  the soliton central frequency and $v_g$ the soliton group velocity. The nonlinear phase shift is small and is usually neglected. The left hand side of equation (\ref{eq:one}) is called integrated dispersion $\beta_{int}$, and can be rewritten as a Taylor expansion leading to:

\begin{equation}
\beta_{int} =  \sum_{k \geq 2} \frac{(\omega - \omega_s)^2}{k!} \left[ \frac{\partial^k \beta}{\partial \omega^k} \right]_{\omega = \omega_s} \simeq 0
\label{eq:two}
\end{equation}

In order to generate DW, the first negative term of equation  (\ref{eq:two}) has thus to be compensated by higher order dispersion terms, while the bandwidth of the soliton at the compression point has to overlap this phase matching region. A first advantage of Si\textsubscript{3}N\textsubscript{4} waveguides thus lies in their very weak Raman response \cite{Karpov2016}. In fact, SCG dynamics usually shows a continuous red shift of the pump laser due to soliton intrapulse Raman scattering <cite{Dudley2010}. According to equation (\ref{eq:one}), this effect would change the phase-matching points along the waveguide, leading to multiple dispersive waves at different wavelengths, for long enough propagation lengths. 
The waveguide dispersion thus determines the location of the phase matching points. It has been shown that the number of phase matched dispersive waves corresponds to the number of zero dispersion wavelengths (ZDW) \cite{Roy2009a,Roy2009}. The material dispersion of Si\textsubscript{3}N\textsubscript{4} can easily lead to a first ZDW point in the near-IR for the fundamental waveguide mode. However the large anomalous material dispersion in the mid-IR has to be compensated by waveguide dispersion in order to reach a second ZDW at longer wavelengths. This poses two challenges for efficient generation of mid-IR DW: first, unwanted visible DW is typically generated when pumping with fiber lasers due to the first ZDW point \cite{Karpov2016}, second, the weak mid-IR mode confinement greatly increases absorption by the silica cladding which surrounds Si\textsubscript{3}N\textsubscript{4}.

\subsection{Waveguide design and simulations}

In order to overcome these limitations, we used large cross-section waveguides, fabricated by conformal deposition of Si\textsubscript{3}N\textsubscript{4} in low pressure chemical vapor deposition \cite{Pfeiffer2016}. The Si\textsubscript{3}N\textsubscript{4} channel waveguides are then buried in SiO\textsubscript{2}.  The waveguide thickness can reaches 2200 nm while, using standard fabrication process, it is limited at 1 $\mu$m.  We combine both mid-IR mode confinement and dispersion engineering, such that the waveguide could support both the pump pulse solitonic regime and Mid-IR dispersive wave generation. 

We retrieve the waveguide GVD knowing the dispersion relation of the refractive index of Si\textsubscript{3}N\textsubscript{4} and computing the waveguide fundamental mode dispersion. In Fig.1a we show that the second ZDW of such thick waveguides can be further red-shifted compared to standard Si\textsubscript{3}N\textsubscript{4} waveguides \cite{Grassani2016}, leading to the generation of DW deeper in the mid-IR for the same pump wavelength. At the same time, the large cross-section waveguide can significantly reduce mid-IR absorption in the silica cladding through improved mode confinement. This can be seen in Fig.1b, where we computed the absorption losses $\alpha$ by including the imaginary part of the refractive index of silica \cite{Kitamura2007} in our numerical simulations. High-order dispersion terms also affect the amount of power transfer to the DW \cite{Akhmediev1995,Roy2009a,Roy2009}. Qualitatively, even order dispersion terms lead to two dispersive waves with symmetric frequency detuning and intensity, while odd order terms break this symmetry favoring blue or red-shifted DW generation in the case they are positive or negative valued, respectively \cite{Roy2009}. In order to increase the mid-IR DW generation efficiency, and at the same time reduce the visible one, the pump should be positioned beyond 1.9 $\mu$m. From our numerical simulations this allows to have the third, fifth and seventh order dispersion negative valued. The amount of integrated dispersion separating the pump to the DW phase matched wavelengths illustrates clearly the symmetry breaking induced by appropriately positioning the pump wavelength. As seen in Fig.1c, a 2090 nm pump in our large cross-section waveguides generates a mid-IR DW in the same wavelength range as the one obtained by using standard waveguides with a 1550 nm pump, as we previously reported \cite{Grassani2016}. However, 2090 nm pumping in the large waveguide leads to a much lower mid-IR dispersion barrier, and a much higher visible one, clearly favoring mid-IR power transfer. 

Finally, the soliton number $N$ does not only influences the spectral extend of the compression point, but also its position $l_c$ along the waveguide as $l_c = L_D/N$ \cite{Dudley2010}. Thus, for a given laser power, pulse duration and waveguide nonlinearity, $\mid <beta_2 \mid$ has to be small enough to allow for a sufficient high soliton number ($N^2 \simeq 1/\mid \beta_2 \mid$) to cover the DW generation phase matching region, but large enough to allow for soliton compression, and thus DW generation, before the end of the waveguide ($l_c \simeq \mid \beta_2 \mid^{-1/2}$).

\begin{figure}
\includegraphics[width=8.5cm]{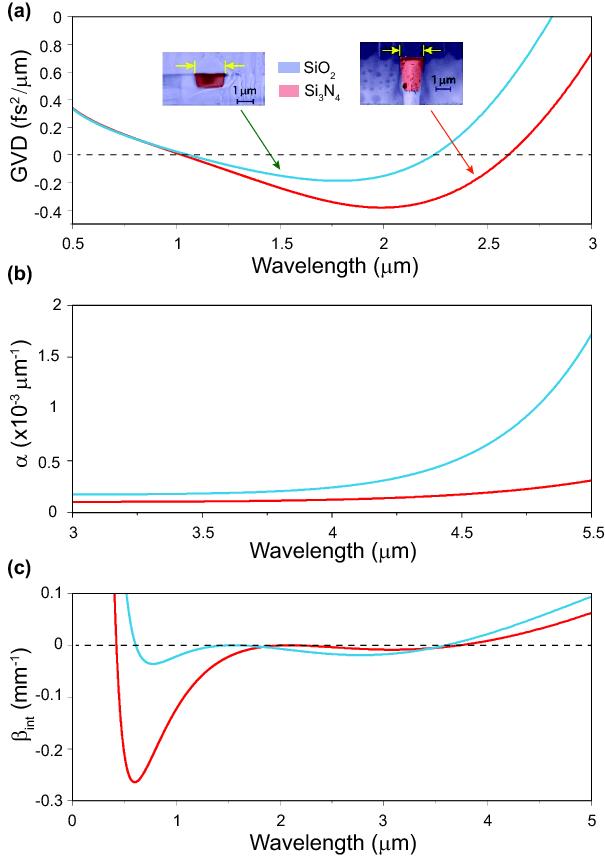}% Here is how to import EPS art
\caption{\label{Figure1} Comparison between standard (blue) and large cross section (red) silicon nitride (Si\textsubscript{3}N\textsubscript{4}) waveguides. (a) Group Velocity Dispersion (GVD) for the TE fundamental mode of a standard waveguide with cross section 870 x 1700 nm\textsuperscript{2} and the TM fundamental mode of a large cross section waveguide with dimension 2177 x 1150 nmtextsuperscript{2}. The insets show Scanning Electron Microscope (SEM) images of the waveguide cross-section. (b) Attenuation coefficient $\alpha$ as a function of wavelength for standard and large cross section waveguide. (c) Integrated dispersion as a function of the wavelength for the standard waveguide pumped at 1560 nm and the large cross section one pumped at 2090 nm. The dispersive wave (DW) phase matching points lie in the same region for both configuration, but the much lower mid-IR dispersion barrier, and the much larger visible one, favors mid-IR DW in the large cross section waveguide.}
\end{figure}

\subsection{Experimental set-up}

\begin{figure*}
\includegraphics[width=17cm]{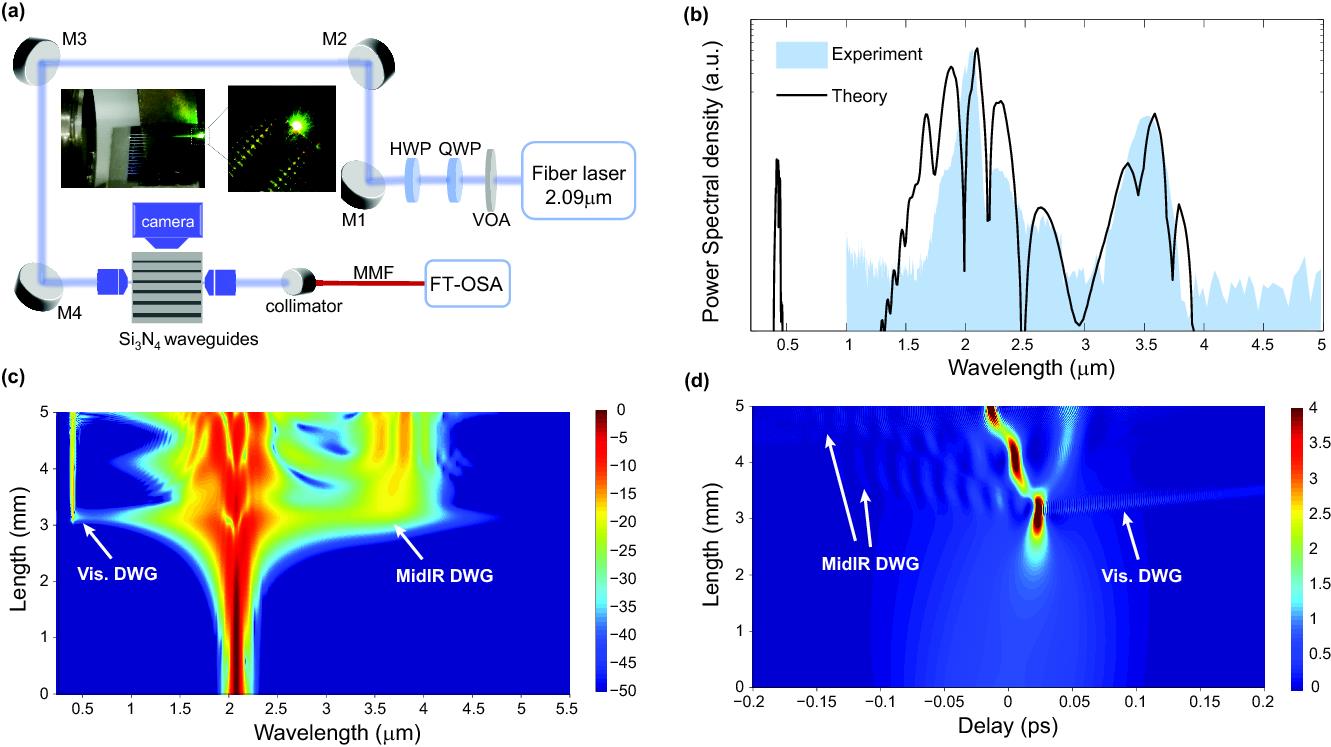}% Here is how to import EPS art
\caption{\label{Figure2}Experimental set-up and dispersive wave generation. (a) Experimental set-up. HWP: half wave plate; QWP: quarter wave plate; M: mirror; MMF: multimode fiber. (b) Experimental spectrum (shaded region) and theoretical simulation (solid line) retrieved in the FT-OSA for 13.6 mW average power coupled in the 1100 nm width waveguide. Spectral (c) and temporal (d) pump pulse evolution (dB scale) over the waveguide length for an input pulse of 110 fs at 2090 nm, with frequency chirp C = -1000 fs\textsuperscript{2}/2$\pi$.}
\end{figure*}

The experimental set-up is detailed in Fig.2a. The pump source is a commercial, turnkey soliton self-frequency shifted Thulium-doped fiber mode locked laser (NOVAE Brevity λ+,) with pulse duration at full width at half maximum of $\sim$ 78 fs and repetition rate of 19 MHz. Polarization management on the beam is carried out through a quarter wave plate and a half wave plate, while the output power is controlled by a variable optical attenuator. Light is coupled in and out of the waveguide’s inverse taper mode converters through two identical aspheric black diamond lenses. At the device output, the collimated light is focused by means of a parabolic mirror onto a fluoride multimode fiber (MMF) (Thorlabs MF12L2) and the spectra are recorded with a Fourier Transform Optical Spectrum Analyzer (FT-OSA) spanning the 1 - 5 $\mu$m range (Thorlabs OSA205C). In order to estimate the actual DW power and CE, we placed a long pass filter with cut-on wavelength at 2.5 $\mu$m at the waveguide output, before the MMF. The DW power at the chip output was obtained by integrating the spectrum within the DW band and correcting this value by the insertion losses coming from the spectrometer, the collimator and the MMF. We estimated the insertion losses coming from the spectrometer by calibrating the measured FT-OSA power values with a known signal, namely the pump power which can be characterized by both the FT-OSA and a semiconductor InGaAs photodiode (Thorlabs S148C). Finally, the on-chip DW power was calculated considering the waveguide output coupling losses through the black diamond lens.

The top view of the device is acquired with a microscope objective which projects the image on a visible camera. The waveguides are straight with a 5 mm length including  including the two inverse taper sections. We investigated waveguides with three different nominal widths: 1050 nm, 1100 nm and 1175 nm. The waveguide thickness also slightly increases with larger widths, ranging from 2.12 $\mu$m to 2.19 $\mu$m. The total coupling losses are estimated to be around 12 dB and are assumed to be equally distributed among input and output, while propagation losses are about 0.2 dB/cm.

\section{Results and discussion}

\begin{figure*}
\includegraphics[width=17cm]{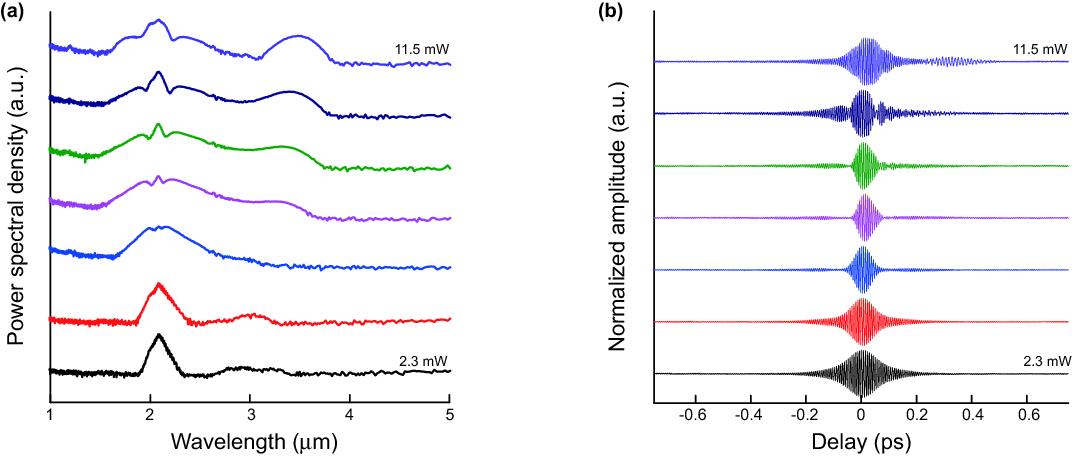}% Here is how to import EPS art
\caption{\label{Figure3}Dispersive wave evolution as a function of the pump power. (a) Power density spectra (dB scale) and (b) field autocorrelation traces recorded at the output of the FT-OSA for coupled average pump power of 2.3, 3.6, 5.8, 7.2, 8.1, 9.1, 11.5 mW.}
\end{figure*}

The experimentally recorded output spectrum from the 1100 nm width waveguide for a coupled average power of 13.6 mW, corresponding to about 6.8 kW peak and coupled pulse energy of 0.75 nJ is presented in Fig.2b. A clear DW is observed at the expected phase matching position point, around 3.5 $\mu$m, when the pump laser is coupled onto the fundamental transverse magnetic polarization mode of the waveguide. In order to examine the DW generation process taking place in our waveguides, we performed numerical simulations based on the nonlinear Schrödinger equation. We consider an input $sech^2$ pulse with a frequency chirp of -1000 fs\textsuperscript{2}/2$\pi$, leading to pulse duration of about 110 fs. The laser temporal pulse broadening mainly comes from the wave-plates and the input objective. We also set the Raman fraction to zero. In Fig.2b, the obtained simulated spectrum reproduces well the experimental one after a propagation of 4.2 mm, which matches the length of the straight waveguide section, without considering the tapered regions. It also shows a weaker visible DW around 500 nm which cannot be detected by the FT-OSA, but justifies the observed green light scattered out of the chip (see Fig.2a). Looking at the simulated pulse evolution over the waveguide length (Fig.2c), we notice how both the visible and mid-IR dispersive waves are generated at the soliton compression point, which takes place at around $l_c$ = 3.5 mm. This is qualitatively confirmed by the appearance of green light scattered out of the chip, at the expected compression point position (Fig2a). Moreover, Fig.2c shows how, after the first compression point, additional spectral broadening points occur, separated by a much shorter distance, before the end of the waveguide. This behavior can be understood looking at the simulated temporal evolution in Fig.2d. In principle, soliton-self compression periodically occurs along the waveguide every soliton length ($L_s = \pi L_D/2$). Just after the first compression point, the propagating pulse separates in two symmetrical pulses in a process known as the soliton splitting effect \cite{Agrawal2013}. In practice, because of high order nonlinear effects which include HOD, self-steepening, DW emission and recoil, one of the pulses has more energy than the other and can undergo sufficient broadening to once again overlap the DW phase matching point. The soliton self-compression process can thus repeat, leading to multiple generation of DW, as long as the new pulse has enough energy to broaden and overlap with the DW phase-matching region. In fact, it has to be noticed that all these pulses are centered around the original pump wavelength, such that the DW phase matching wavelength remains identical all-over the length of the waveguide.

\begin{figure*}
\includegraphics[width=17cm]{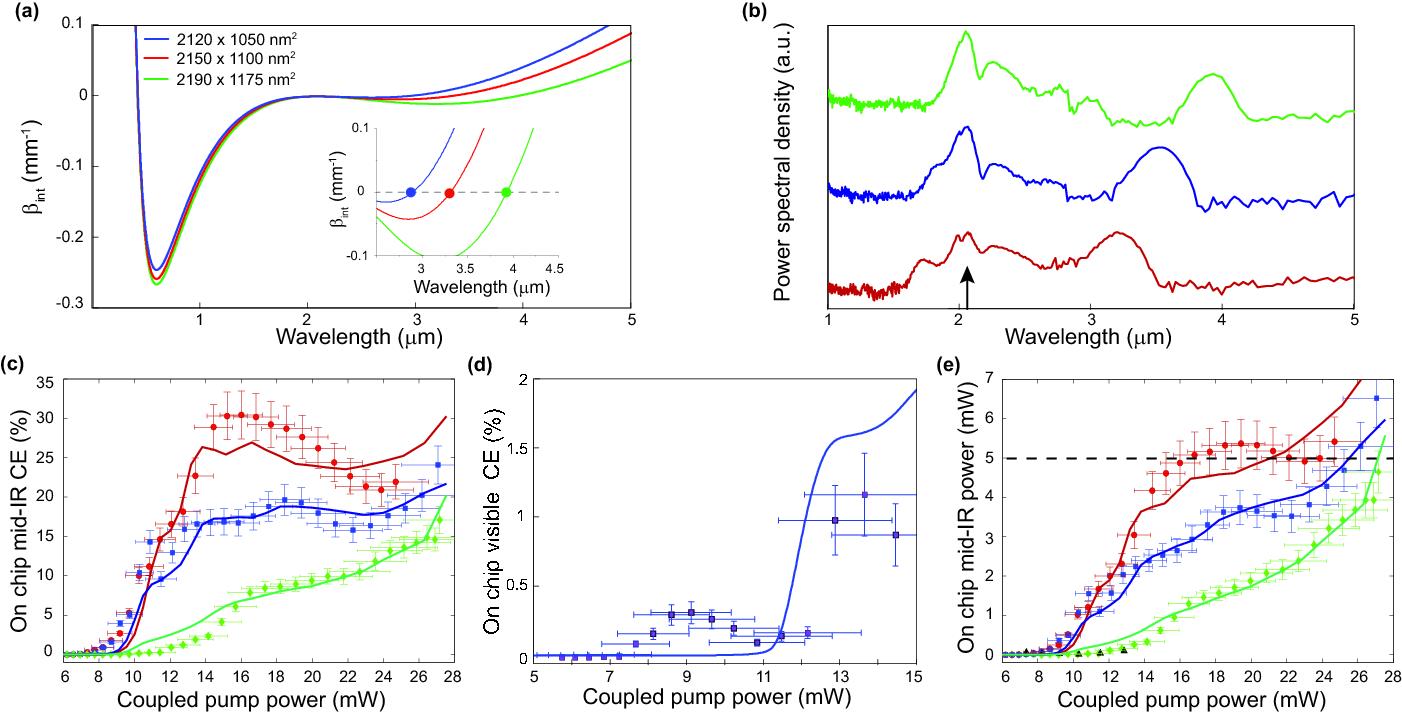}% Here is how to import EPS art
\caption{\label{Figure4}Dispersive wave generation efficiency. (a) Overall integrated dispersion for the waveguides under test:  2120 x 1050 nm\textsuperscript{2} (blue), 2150 x 1100 nm\textsuperscript{2} (red) and 2190 x 1175 nm\textsuperscript{2} (green). The inset shows the mid-IR phase matching point. (b) Experimental spectra recorded at an estimated average coupled power of 13.6 mW, for the four different waveguides under test (same color convention as for fig.4a).  Arrow indicates pump wavelength. (c) Mid-IR dispersive wave generation on-chip conversion efficiency (CE) as a function of coupled pump power. Points are experimental values while the error comes from the ±0.5 dBm and ±0.25 dBm uncertainty in the power of the dispersive wave and the coupled pump, respectively, lines are simulations. (d) Visible dispersive wave generation efficiency as a function of the pump power for the waveguide with cross section 2120 x 1050 nm\textsuperscript{2}. Scattered points represent the experimental data, line is simulation. The error bars on the y axis take into account ± 1 dB uncertainty on the measurement of the visible dispersive wave power, while error bars on the x axis take into account ±0.25 dB uncertainty on the coupled power. (e) On-chip Mid-IR dispersive wave power estimated from the measured spectra as detailed in Section the Experimental setup section 3. The uncertainties in the dispersive wave and pump power are the same as for Fig.4c. Lines are simulations. 5mW of on chip DW power is reached at 3220 nm with 16 mW of pump, at 3530 nm with 25 mW and 3950 nm with 28 mW.}
\end{figure*}

In Fig.3a we report the measured spectral evolution for different input coupled pump power for the 1100 nm wide waveguide. The onset of DW generation is detected starting from an average coupled pump power of approximately 7.2 mW. The DW then grows with increasing coupled pump power while, as expected, no substantial change in the soliton central frequency nor in the mid-IR DW wavelength is detected.  The emission of the dispersive wave can also be clearly seen from the field autocorrelations reported in Fig.3b. As the dispersive wave propagates with a different group velocity with respect to the main pulse, two additional interference patterns at a different time delay are created.   

From the simulated integrated dispersion shown in Fig.4a for the three waveguides under test, it can be noticed how the mid-IR DW phase matching point continuously shifts toward longer wavelengths by increasing the waveguide width. Importantly, in all waveguides, the integrated dispersions show how the pump wavelength at 2090 nm significantly favors mid-IR DW with respect to the visible one. The experimental spectra shown in Fig.4b confirm the theoretical predicted phase matched wavelength for the generation of the mid-IR DW, with the central wavelength red shifted by increasing the waveguide width. We measure dispersive waves at 3220 nm, 3530 nm and 3950 nm for the 1050 nm, 1100 nm and 1175 nm waveguide width, respectively, resulting in the entire coverage of the 3 to 4 $\mu$m spectral region. The spectral width of the generated dispersive waves slightly decreases by increasing the central wavelength: we measure a bandwidth of 6.9 THz, 6.3 THz and 4.2 THz for the three waveguides, respectively. The spectra of Fig.4b also show that, apart for some pump broadening due to self-phase modulation, most of the converted pump energy lies in the DW bandwidth.

 Fig.4c shows the experimental points and the numerical simulation of the on chip CE for all the tested waveguides as a function of the coupled pump power. To better compare with the theoretical results, we include a nonlinear contribution in the estimate of the coupled pump power. In fact, increasing the pumping level, both multiphoton absorption processes and free carrier absorption, mainly due to the absorption of the generated visible DW, come to play a role. From the data related to the 1050 nm and 1100 nm, it is quite clear how, overall, the CE goes through three regimes. First, at low power, where the soliton is still not compressed and no DWs are generated. Beyond a certain pump power threshold, soliton compression occurs before the end of the waveguide, leading to the emission of DWs. The efficiency of this process rapidly increases with pump power, reaching a record value of more than 30\% for the 1050 nm width waveguide. This comes from the emergence of multiple compression points within the waveguide length, which can reinforce the DW, as detailed in the section of Methods and Materials. After shedding energy multiple times to the DW, the soliton power eventually drops, limiting once again the spectral broadening. We can thus identify a third power level (saturation power), corresponding to the minimum value which includes all the compression points leading to the generation of DW before the end of the waveguide. After the saturation power, the CE has a much lower slope and can be consider almost constant. We calculate a soliton number $N \simeq$ 7 for the maximum coupled pump power in our experiment. Further increasing the pump power to $N >$ 10 would lead to noise seeded broadening, where power from the pump would no more be converted to a preferential spectral region and soliton dynamics does not apply any more. 
Increasing the width of the waveguide, the saturation seems to happen for higher pumping levels. This is expected as, looking at Fig.4a, the mid-IR dispersion barrier increases for larger waveguides reducing the power transfer from the pump to the DW. The 1175 nm width waveguide features a slower increase in its efficiencies and there is no a clear change in its CE slope after the power threshold, indicating that probably we did not reach saturation within the employed pump range. Nevertheless, in this waveguide, we reach close to 20\% of CE for a DW centered at 3950 nm. Interestingly, the CE curves in Fig.4c are not strictly monotonic, but present oscillations. We attribute this to the multiple compression points appearing in the soliton dynamics for sufficiently high power levels (see Fig.2c and d). In fact, when a compression point happens at the waveguide end, we evaluate, both experimentally and numerically, the fraction of the temporally compressed soliton power which overlaps the DW region. However, this power will not be completely transferred to the DW afterward. Qualitatively, this explains why, after every compression point, the efficiency of the process decreases. Moreover, as observed, this effect is expected to be more pronounced when DWs are generated closer to the pump, where also the power level in the temporally compressed soliton spectrum is larger. We can observe roughly three oscillations in the CE for the 1100 nm in Fig.4c, which are in agreement with the number of compression points observed in Fig.2c for 13.6 mW input power. 

We also estimated the visible DW contribution by changing the output objective with a silica lens and the long pass filter with a short pass one with cut-off at 700 nm. The transmitted light was then measured directly on a Si detector (Thorlabs S120C). This configuration had much larger output losses than the previous one leading to a larger uncertainty on the measurement. The results for the on-chip CE in the 1050 nm waveguide are shown in Fig.4d together with the numerical prediction. We indeed confirm a much lower visible DW generation efficiency (less than 1.5\%) and an overall good agreement with theory.
Although larger waveguides have lower efficiencies, roughly the same DW power is achieved in the 1050 nm, 1100 nm and 1117 nm width waveguides. This can be seen in Fig. 4e where we report numerical and experimental data of the Mid-IR DW power as a function of the coupled pump power. Indeed, in the employed pump power range, all waveguides can generate mid-IR DW up to 5 mW on-chip. The required pump power increases with the wavelength reach. Overall, this means that more than 1 mW can be retrieved at the waveguides output, covering the entire 3.2 $\mu$m to 4 $\mu$m.

These values represent a significant improvement both in terms of spectral coverage and efficiency, compared to previous mid-IR SCG \cite{Guo2018}.  Notably, the mid-IR output power we obtained at 4 $\mu$m is comparable to the state-of-the-art mid-IR sources for dual comb spectroscopy, based on DFG in PPLN waveguides \cite{Ycas2018}, pumped with a similar fiber frequency comb. Moreover, here we can obtain such power levels with roughly ten times less average input power.

\section{Conclusions}

In conclusions, we showed that the generation efficiency of mid-IR DW from a commercial femtosecond SWIR fiber laser can be greatly enhanced, up to about 30\%, by proper dispersion engineering of Si\textsubscript{3}N\textsubscript{4} waveguides. In fact, state-of-the art fabrication techniques allows a careful waveguide design which, in combination with a proper choice of the pump parameters, significantly favor the generation of mid-IR DW and reduces the visible one, allowing to report the highest efficiency measured to date for mid-IR DW generation. The central wavelength of the dispersive waves can be set at will with lithographical precision inside the first mid-IR transparency window by simply changing the waveguide width. Although the efficiency decreases with increasing the DW wavelength, we were able to obtain, at the chip output, a maximum average output power of more than 1 mW spanning roughly the entire 3 - 4 $\mu$m region. 

The presented approach can lead to a very efficient, compact and easy to use device for coherent mid-IR light generation. In fact it benefits from photonic integration in its main parts: a chip scale nonlinear stage compatible with planar fabrication techniques and a silica fiber-based pump source. The device can reach the 4 micron region, which hosts the signature of important greenhouse gases, with a power level sufficient for spectroscopy application \cite{Ycas2018,Schliesser2005},  bridging thus the gap between fiber sources and quantum cascade lasers, which are the workhorse of mid-infrared spectroscopy devices. 
Such result can therefore provide a suitable alternative to microresonators \cite{Luke2015,Griffith2015a,Wang2013,Yu2016} to generate mid-IR frequency combs on a chip, when lower repetition rates are required. Moreover, soliton induced SCG in integrated photonic platforms has been recently demonstrated to coherently broaden the spectrum of optical frequency combs by more than one octave, allowing their stabilization in a $f$ to $2f$ scheme \cite{Johnson2015a,Klenner2016b,Mayer2015a}. Therefore, with the possibility to combine both $\chi^{(2)}$ and $\chi^{(3)}$ nonlinearities in  Si\textsubscript{3}N\textsubscript{4} \cite{Billat2017,Porcel2017}, full on-chip stabilization of the mid-IR comb generated in these waveguides can also be considered.

\begin{acknowledgments}
D.G., E.T. and C.-S.B. acknowledge support by the European Research Council (ERC) under grant agreement ERC- 2012- StG 306630-MATISSE. C.H., H.G. and T.J.K. acknowledge support by contract W31P4Q-16-1-0002 (SCOUT) from the Defense Advanced Research Projects Agency (DARPA), Defense Sciences Office (DSO), and support by the Swiss National Science Foundation under grant agreement No. 161573; H.G. acknowledges funding from the European Union’s Horizon 2020 research and innovation programme under Marie Sklodowska-Curie IF grant (No. 709249). 
\end{acknowledgments} 

%\appendix*
\section*{Conflict of interests} 
The authors declare no conflict of interests.
%\end{section}

\section*{Author contributions}
D.G. performed the simulations, and D.G. and E.T. performed the experiments under the supervision of C.S.B. . H.G. and C.H. conceived the design of large-cross-section Si3N4 waveguides under the supervision of T.J.K. C.H. fabricated the large-cross-section waveguides. All authors discussed the data. D.G. and E.T. wrote the manuscript with input from others. C.S.B. supervised the project.  
%\end{Author contributions}

\newpage %Just because of unusual number of tables stacked at end

\bibliography{library}% Produces the bibliography via BibTeX.
\end{document}